\documentclass[twocolumn,prl,superscriptaddress,showpacs]{revtex4}
\usepackage{epsfig}
\begin{document}
\title{Disentangling the exchange coupling of
entangled donors in the Si quantum computer architecture}
\author{Belita Koiller}
\affiliation{Condensed Matter Theory Center, Department of Physics, 
University of Maryland, College Park, MD 20742-4111}
\affiliation{Instituto de F{\'\i}sica, Universidade Federal do 
Rio de Janeiro, 21945, Rio de Janeiro, Brazil}
\author{Xuedong Hu}
\affiliation{Condensed Matter Theory Center, Department of Physics,
University of Maryland, College Park, MD 20742-4111}
\affiliation{Department of Physics, University at Buffalo, SUNY,
Buffalo, NY 14260-1500}
\altaffiliation[Present address.]{} 
\author{H.D. Drew}
\affiliation{Condensed Matter Theory Center, Department of Physics,
University of Maryland, College Park, MD 20742-4111}
\author{S. Das Sarma}
\affiliation{Condensed Matter Theory Center, Department of Physics,
University of Maryland, College Park, MD 20742-4111}
\date{\today}
\begin{abstract}
We develop a theory for micro-Raman scattering by single and coupled 
two-donor states in silicon.  We find the Raman spectra to have significant
dependence on the donor exchange
splitting and the relative spatial positions of the two donor sites.  In
particular, we establish a strong correlation between the temperature
dependence of the Raman peak intensity and the inter-donor exchange 
coupling.
Micro-Raman scattering can therefore potentially become a powerful tool to
measure interqubit coupling in the development of a Si quantum computer
architecture.
\end{abstract}
\maketitle  

Quantum computing has attracted wide-spread interest in recent years
\cite{Reviews}.
Among many proposed schemes for quantum computer (QC) hardware, a
particularly promising architecture is the nuclear spin based Si QC
based on donor arrays \cite{Kane},
which has prompted a great deal of activities to study
both theoretically and experimentally the fabrication, control, and
measurement of such Si-based devices
\cite{Vrijen,Kane2,Privman,Levy,Clark,KHD1,KHD2,Tahan,Roger}.
Currently the fabrication of donor arrays in Si has been pursued
using two complementary approaches.  The ``bottom-up'' approach uses low
temperature MBE and STM to build the array atom by atom \cite{Clark}, while
the ``top-down'' approach uses ion-implantation \cite{Implantation}.
In both approaches, it is crucial to know the precise
positions of the donors buried inside the host Si lattice,
and to measure the exchange coupling between neighboring donors.  

The most important (and daunting from the fabrication perspective
of building a Si QC) physical quantity underlying the Si QC architecture
is the exchange entanglement of the neighboring donor sites -- this
exchange coupling must be large enough for two-qubit operations to be
possible \cite{Kane,Vrijen,Kane2,Privman,Levy}.  Right now, however,
there is no proposed method for
directly estimating the exchange entanglement in Si, and the
experimental fabrication of Si QC architecture is essentially based on
the hope of appreciable entanglement in the fabricated devices.   In this
Letter we not only propose a novel and promising quantitative method for
obtaining the exchange coupling in Si donor states based on the easily
available technique of electronic micro-Raman spectroscopy 
\cite{microR,Infrared},
but also demonstrate through quantitative calculations the feasibility of
our proposed technique as a powerful diagnostic tool for studying buried
donor positioning and exchange coupling measurement.
Electronic Raman spectroscopy has been widely used
to study electronic transitions for which one-photon
absorption \cite{absorption} is forbidden.  Furthermore, micro-Raman
spectroscopy, with the laser spot size in the order of $\mu$m, has the
required spatial resolution to provide valuable information on the
environment of single donors, pair orientation, and the
strength of donor-pair coupling and electronic entanglement.  In the
following, we develop a theory for micro-Raman scattering from donor 
electrons in Si, focusing on the situation where
photon polarization and sample temperature can be precisely tuned.  
We show that the electronic micro-Raman spectra have important signatures
of donor locations and exchange coupling.  We also discuss specific
experimental conditions when single donor (or single pair) detection
might be possible with this technique.

In Raman scattering (i.e. inelastic light scattering)
experiments, an incident laser photon of wave
vector $\vec k_L$, frequency $\omega_L$ and polarization $\vec\eta_L$
scatters into a Raman photon with $\vec k_R$, $\omega_R$, $\vec
\eta_R$, while the system with which the radiation interacts
undergoes a transition from an initial state $|0\rangle$ of energy
$E_0$ to a final state $|f\rangle$ of energy $E_f$.
The differential Stokes-Raman scattering cross section is given by
\cite{klein}
\begin{equation}
\frac{d^2 \sigma}{d\Omega_R d\omega_R}= \left(\frac{e}{mc}\right)^4\, 
\frac{\omega_R}{\omega_L}\sum_f |M|^2 \delta(\omega-\omega_{f0})\, ,
\label{eq:cross}
\end{equation}
where $\omega_{f0} = (E_f - E_0)/\hbar$ and $\omega = \omega_L -
\omega_R$ is the Raman shift.  The matrix element $M$ is written
as \cite{klein}
\begin{equation}
M = \sum_{\mu,\nu = x,y,z} \eta_L^\mu \eta_R^\nu
\left(\frac{\langle f|p_\mu p_\nu |0\rangle}{E_G - \hbar \omega_R} 
+ \frac{\langle f|p_\mu p_\nu |0\rangle}{E_G + \hbar \omega_L}
\right)\, ,  
\label{eq:symmetry}
\end{equation}
where $E_G$ is the direct band gap at the conduction band minimum and
$\vec p$ is the momentum operator.  In Si, the conduction
band has six degenerate minima.  The lowest donor-electron
energy states are a ground-state singlet $1S(A_1)$, and excited
triplet $1S(T_2)$ and doublet $1S(E)$ components, consistent with the
$T_d$ site symmetry of the substitutional donor \cite{Kohn}.  The
notation $1S(j)$ refers to a $j$-symmetry state obtained from 
1S-like hydrogenic envelopes, which can be written as \cite{KHD2}
$\psi_j ({\bf r}) = \sum_{\mu = 1}^6 \alpha_\mu^{(j)} F_{\mu} ({\bf r})
\phi_\mu(\bf r)$,
where $\phi_\mu({\bf r}) = u_{\mu}({\bf r})e^{i {\bf k}_{\mu}\cdot 
{\bf r}}$ are Bloch wavefunctions, $F_{\mu} ({\bf r})$ are the
hydrogenic envelopes, $\alpha_\mu^{(j)}$
characterize the point group representation, and $\mu$ 
runs over the six conduction band minima ${\bf k}_{\mu}$.  
Taking this multivalley feature into consideration,
Eq.~(\ref{eq:cross}) can be simplified
\cite{klein}:
\begin{eqnarray}
\frac{d^2 \sigma}{d\Omega_R d\omega_R} & = & \left( \frac{e^2}{mc^2}
\right)^2\, \frac{\omega_R}{\omega_L} {\cal R}^2
(m_{\parallel}^{-1} - m_{\perp}^{-1})^2 \nonumber \\
& & \hspace*{-0.6in} \times \left| \sum_{\mu=1}^6 \alpha_{\mu}^{(f)*}
\alpha_{\mu}^{(0)} \, (\vec{\eta}_L \cdot \hat{\eta}_{\mu}) 
(\vec{\eta}_R \cdot \hat{\eta}_{\mu})
\right|^2 \delta(\omega-\omega_{f0})\,, 
\label{eq:cross_f} 
\end{eqnarray} 
where ${\cal R}$ is the resonance enhancement factor which, 
for $\hbar(\omega_L-\omega_R)\ll E_G$, can be expressed as
${\cal R} =E_G^2/[E_G^2 - (\hbar \omega_L)^2]$, 
$m_{\parallel}$ and $m_{\perp}$ are the Si conduction band effective masses
in units of the free electron mass $m$, and $\hat{\eta}_{\mu}$ is
the unit vector along the axis of the $\mu$th valley.

Electronic Raman scattering in Si:P is dominated by the valley-orbit
transition of $1S(A_1) \to 1S(E)$ states for single donors (the $1S(A_1)
\to 1S(T_2)$ transition is not Raman-active) \cite{klein}.
The shift and width of
the single donor Raman peak can provide donor environment information,
since the donor states are sensitive to
the local strain fields, the local electric fields from ionized
acceptor-donor pairs or surface charge for donors near the
surface, and interactions with other neutral donors and
acceptors.  

To have single donor sensitivity, a micro-Raman experiment needs
sufficiently large Raman cross section, which is possible through the
resonant Raman effect.
The Raman cross section can be estimated neglecting the
excitonic effects \cite{excitonic}.
In this case the
divergence in the resonance enhancement factor $\cal{R}$ is limited by
the energy level dispersion in the valence band, since the valence bands 
are not extremal at the conduction band minima.
As a result, ${\cal R}$ can take on
a maximum value of ${\cal R}^2 \approx 1000$,
leading to a Raman cross section $d\sigma /d\Omega_R \approx
2\times 10^{22}$ cm$^2$/sr.  For example, for a laser power
of 1 mW the photon flux is $1.6 \times 10^{23}$ photons/s $\cdot$
cm$^2$.  This leads to a count rate of about 35 s$^{-1}$ per donor,
small but workable.  In addition, at
$T=10$K and for the incident laser power of 1 mW in a 1 $\mu$m spot, 
the sample temperature rise is estimated to be only 1 K.
The spatial resolution needed for micro-Raman experiments is also
facilitated by the fact that the resonance condition occurs in the UV
($\sim 3.8$ eV) where the penetration depth of the radiation is small
($\sim$ 10 nm).  For a sample with a donor density of $10^{-16}$
cm$^{-3}$ and a focal spot size $\sim$ 1 $\mu$m there are $\sim$ 100
donors in the field of view.  At this density
the mean number of donor pairs with separations within 10 nm is about 4.

If resonance enhancement allows sufficiently large signal-to-noise
ratio for micro-Raman scattering, an immediate question is what 
information of the phosphorus donors the Raman spectra carry,
and how they are related to donor positions and interdonor coupling.
Here we focus on the Raman spectra of a
pair of donors at ${\bf R}_A$ and ${\bf R}_B$ (relative position
${\bf R} = {\bf R}_A - {\bf R}_B$), namely, when the $|0\rangle$ and
$|f\rangle$ states of Eq.~(\ref{eq:symmetry}) are two-donor states.
It is thus necessary to study the
composition and energy spectra of the Raman-active donor-pair
states, which is the theory we develop here.

\begin{figure}
\centerline{
\epsfxsize=3.4in
\epsfbox{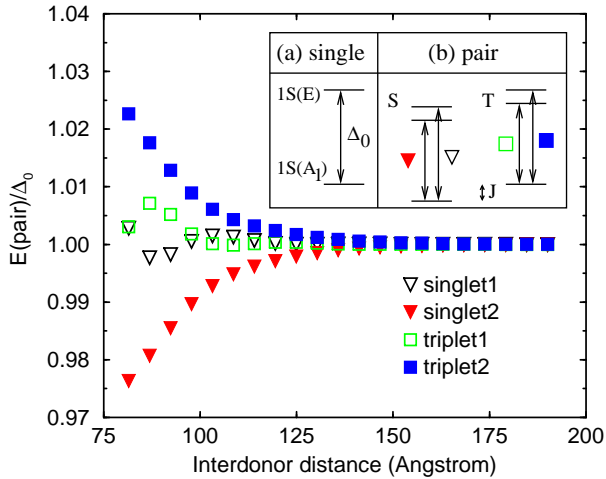}}
\vspace*{0.1in}
\protect\caption[Raman shifts versus distance]{
Raman shifts of the four active transitions for a pair of
substitutional donors placed along the [100] direction in Si,
relative to the single impurity value $\Delta_0$.
Different symbols are defined in inset (b) and in the text.
Insets: (a) Energy levels of a single P donor in Si leading to a
Raman shift $\Delta_0=105$ cm$^{-1}$.
(b) Schematic representation of the two-electron states
of a P donor pair that participate in the Raman process.
}
\label{fig-ExR}
\end{figure}

In the limit of distances $|{\bf R}|$ much larger than the effective Bohr
radii, consistent with the Si QC proposals \cite{Kane,Vrijen,Kane2},
the two-particle singlet and triplet states within the $1S$ manifold
may be written in the Heitler-London (HL) approach:   
\begin{eqnarray}
\psi_{\stackrel{s}{t}}^{i,j} ({\bf r}_1, {\bf r}_2)
& = &  \{u[\,|A_1^i,B_2^j\rangle\pm |B_1^j,A_2^i\rangle] \nonumber \\
& & \hspace*{-0.2in}+v[\,|A_1^j,B_2^i\rangle \pm
|B_1^i,A_2^j\rangle]\} \left( |\uparrow \downarrow \rangle \mp
|\downarrow \uparrow \rangle \right) \,,  
\label{eq:wavef} 
\end{eqnarray}
where $|A_1^i,B_2^j\rangle =  \psi_i({\bf r}_1 - {\bf R}_A)
\psi_j({\bf r}_2 - {\bf R}_B)$.  Here $A$ and $B$ refer to the two
nuclei locations, $i$ and $j$ refer to the single donor electronic
states, $1$ and $2$ refer to the two electrons, and only one of the
spin triplet is shown.
Normalization and symmetry define the $u$ and $v$ coefficients.  The two
lowest eigenstates are singlet and triplet states obtained from the
single donor ground state $i=j=A_1$.  Their splitting $J(\bf R)$
defines the Heisenberg exchange coupling between the ground-state
donor electrons [see inset (b) in Fig.~\ref{fig-ExR}], and is a 
direct measure of the degree of control achievable over the
entangled electronic states given in (\ref{eq:wavef}).  When the
initial state $|0\rangle$ is a singlet or a triplet with $i=j=A_1$,
the lowest Raman-active final states $|f\rangle$ are obtained for
$i=A_1$ and $j=E^{(\kappa)}$, where $\kappa = 1,2$ refer to the two
doublet states whose degeneracy has been lifted due to the axial
perturbation of the donor pair. 

There are 2 initial and 8 final states for Raman
scattering from a donor pair in the $1S$ manifold.
The 16 possibilities reduce to 4 Raman active combinations due to the
absence of spin transitions (we
neglect spin-orbit interactions) and the $A \leftrightarrow B$
symmetry of the HL state (\ref{eq:wavef}): Two of type $\{|0\rangle_s ,
|f^{(\kappa)} \rangle_s^+\}$ and two of type $\{|0\rangle_t ,
|f^{(\kappa)} \rangle_t^-\}$, $\kappa = 1,2$.  The superscript
$\pm$ indicates $u=\pm v$ in (\ref{eq:wavef}).

The Raman allowed lines are schematically represented in
the inset (b) of Fig.~\ref{fig-ExR}, and their calculated values for a
substitutional impurity pair positioned along the [100] direction are
given in Fig.~\ref{fig-ExR}, where triangles (squares) label singlet
(triplet) lines, and open (filled) symbols correspond to $\kappa =
1(2)$.  The behavior of the calculated Raman shifts
in Fig.~\ref{fig-ExR} illustrates general
trends obtained for any orientation of $\bf R$, namely the overall
convergence of all lines to the single-impurity value as $|{\bf R}|$
increases, superimposed with an oscillatory behavior (more apparent
for $\kappa = 1$ for this particular direction) due to
intervalley electronic interference of Si conduction
band minima \cite{KHD1,KHD2}.  In general, Raman
transition energies are anisotropic and very sensitive to the relative
positioning of the donor pair. 

The propagation and polarization directions of the laser and 
Raman-scattered light, as well as the system temperature, are
controllable parameters in experiments.  The usual notation to
indicate the polarization scheme is $P = \vec k_L(\vec\eta_L,
\vec\eta_R)\vec k_R$.  Here we have considered the
following: $P_1 = Z(X,X)\bar Z$, $P_2 = Z(Y,Y)\bar Z$, $P_3 =
Z([110],[1\bar 10])\bar Z$ and $P_4 = Z([110],[110])\bar Z$,
where $X$, $Y$, $Z$ are the crystal axis [100], [010] and [001].  The
relative Raman intensity of the different lines in a single spectrum
is strongly dependent on the polarization
components [see Eq.(\ref{eq:symmetry})].  We also consider
temperature effects in the populations of the initial states
$|0\rangle_s$ and $|0\rangle_t$, assumed to be in
thermal equilibrium, i.e.
\begin{equation}  
n_t/n_s = 3 \exp (-J/k_B T)\, .
\label{eq:population} 
\end{equation}
\begin{figure}
\centerline{
\epsfxsize=3.2in
\epsfbox{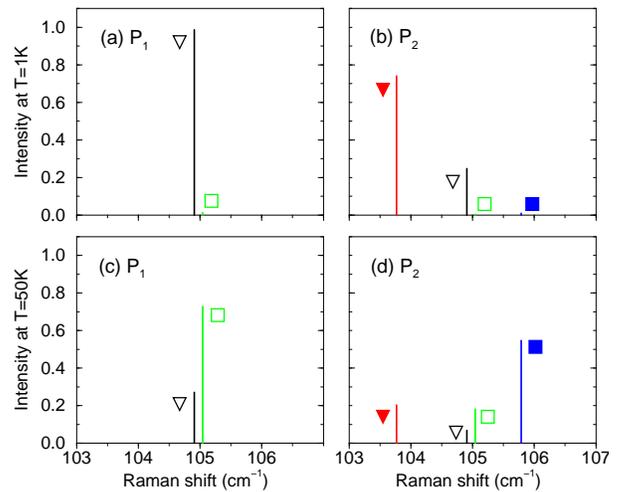}}
\vspace*{0.1in}
\protect\caption[Raman lines]{
Polarization and temperature effects in the Raman spectrum
of an impurity pair separated by $\sim 100$~\AA~ along [100].
Each vertical line represents a $\delta$-function with weight
(line intensity) proportional to its length.
Frames (a) and (c) correspond to polarization $P_1$, parallel to
$\bf R$,  and (b) and (d) to polarization $P_2$, perpendicular
to $\bf R$. Temperatures are as indicated on the left.}
\label{fig-lines}
\end{figure}

In Fig.~\ref{fig-lines} we illustrate the combined polarization and
temperature effects in the Raman spectrum for a pair of
substitutional P donors along the [100] axis separated by $\sim 100$
\AA (18 lattice parameters).  Frames \ref{fig-lines}(a) and 
\ref{fig-lines}(b) are the data for low
temperature ($T=1$K), while frames \ref{fig-lines}(c) and 
\ref{fig-lines}(d) are high temperature
data ($T=50$K).  In addition, frames \ref{fig-lines}(a) and 
\ref{fig-lines}(c) refer
to polarization $P_1$, i.e., along $\bf R$.  Here only two
lines appear, because the $\kappa = 2$ lines (solid symbols in
Fig.~\ref{fig-ExR}) yield $M=0$ for $P_1$.  The spectrum becomes more
interesting for polarization $P_2$, perpendicular to $\bf R$, shown
in frames \ref{fig-lines}(b) and \ref{fig-lines}(d).  Here all four 
lines contribute to the spectrum. 
At low temperatures [frames \ref{fig-lines}(a) and \ref{fig-lines}(b)], 
practically all of the
spectral weight is concentrated in the singlet lines, while for
$T=50$ K [frames \ref{fig-lines}(c) and \ref{fig-lines}(d)] all lines 
contribute, and the triplet
lines are almost 3 times stronger than the corresponding (same
$\kappa$) singlet lines due to the triplet spin degeneracy.
Polarizations $P_3$ and $P_4$ lead to Raman lines qualitatively 
similar to $P_2$, but with reduced total intensity.
We find that Raman intensities are stronger for
polarization configurations such that $\vec\eta_L$ and $\vec\eta_R$
are either parallel or perpendicular to the relative position
vector ${\bf R}$ (such as $P_1$ and $P_2$ in the present case
of ${\bf R}$ along [100]), a feature that can be explored
experimentally to determine the pair orientation.  

\begin{figure}
\centerline{
\epsfxsize=3.4in
\epsfbox{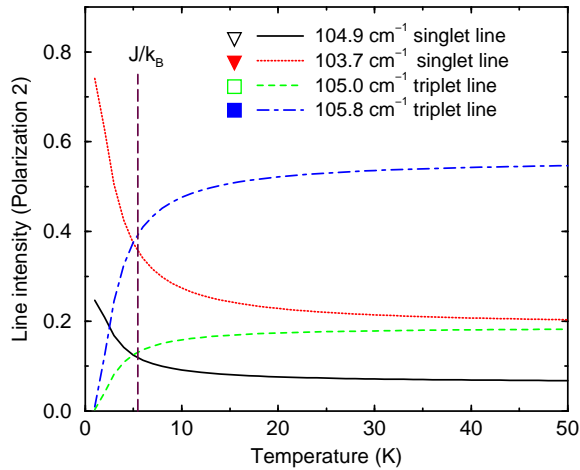}}
\vspace*{0.1in}
\protect\caption[intensity versus temperature]{
Relative intensities of the Raman lines of an impurity
pair for which the exchange coupling is $J=0.47$ meV.
Note that there is a crossover between singlet and triplet
pairs at a temperature close to $J/k_B$, allowing to
extract quantitative information about $J$}
\label{fig-IxT}
\end{figure}

In Fig.~\ref{fig-IxT} we present the individual line intensities for
increasing temperatures.  The intensity of same-$\kappa$-pairs of
singlet-triplet lines undergo a crossover close to the temperature
$J/k_B$, a result easily obtained from Eq.~(\ref{eq:population}).
Thus $J$ may be quantitatively estimated from the temperature
dependence of the Raman line intensity.  For $J \ll k_B T_{\rm min}$,
the lowest experimentally achievable temperature, no effect should be
observed by raising $T$, thus providing an upper bound for the value
of $J$. 
\begin{figure}
\centerline{
\epsfxsize=3.4in
\epsfbox{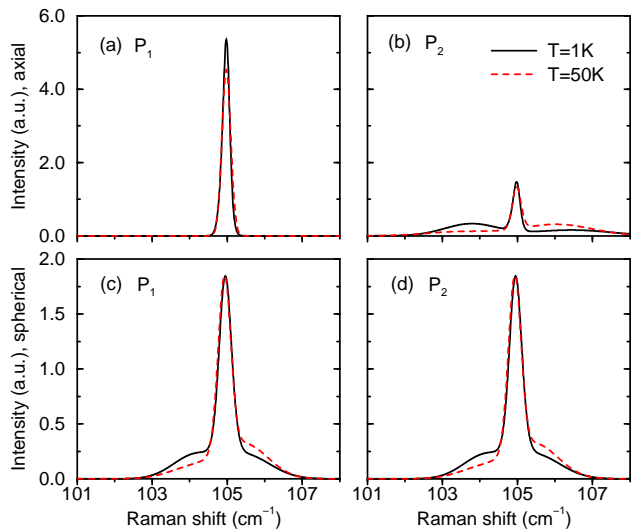}}
\vspace*{0.1in}
\protect\caption[Gaussian fits]{
Raman signature of non-Poissonian distribution of donor pairs.
In panel (a) and (b) the pair distribution has axial symmetry,
while in (c) and (d) the symmetry is spherical.}
\label{fig-distributions}
\end{figure}

We have also investigated the Raman signature of an ensemble of 
donor pairs, which is within the current experimental
capability.  We consider a sample with an {\it ensemble of donor
pairs} of average relative position ${\bf R}_0=100$~\AA along
[100].  For each pair, with a given ${\bf R}_B$, ${\bf R}_A$ varies
over all possible fcc lattice sites within a sphere of radius 10~\AA
centered at ${\bf R}_B+{\bf R}_0$.
The resulting spectra are given in Figs.~\ref{fig-distributions}(a) and
\ref{fig-distributions}(b).  The Raman signature with polarization 
$P_2$ shows an
interesting $T$ dependence: as $T$ increases, the spectral
weight shifts upward in frequency, so that a low-energy shoulder at
$T=1$K becomes a high-energy shoulder at $T=50$K. 
Our results thus show that conventional Raman experiments may be
useful in identifying whether non-Poissonian axial distributions of 
donor pairs were actually achieved.
Figs.~\ref{fig-distributions}(c) and \ref{fig-distributions}(d) 
give results for a
different type of pair distribution, also of average relative
position $|{\bf R}_0|$, but with spherical symmetry: ${\bf R}_A$
varies over all possible fcc lattice sites between two spherical
shells of radii $|{\bf R}_0\pm 10|$~\AA centered at ${\bf R}_B$.
Here the Raman signatures of the two polarizations are essentially
identical, reflecting the sample's isotropic nature.

In this Letter we have proposed and analyzed a powerful and
novel optical technique for disentangling donor exchange coupling
information in the Si QC architecture.  Our proposed technique actually
transcends the specific problem (i.e. Si QC architecture fabrication)
that motivated us--the proposed micro-Raman spectroscopic
technique has the potential of becoming a versatile diagnostic tool in
nanofabrication in general, since it can provide precise spatial
information about relative positioning
and entanglement of neighboring atoms in a host material, 
something that is very hard to do with any
currently existing tools in solid state physics.

We thank LPS, ARDA and NSA for financial support, and Danilo Romero
for useful conversations.  BK also
acknowledges financial support from CNPq (Brazil).


\begin{thebibliography}{}
\bibitem{Infrared} Infrared absorption into optically-active levels
\cite{absorption} can be another useful tool in studying donor
coupling in Si.  However, the lack of any resonant enhancement
severely limits the resolution of such an experiment.
\end{thebibliography}
\end{document}